**The relationship between the number of editorial board members and the scientific output of universities in the chemistry field**

**Xing Wang[1, 2] \***

1 School of Information Management, Shanxi University of Finance & Economics, 696 Wucheng Road, Taiyuan, P. R. China

2 Graduate School of Education, Shanghai Jiao Tong University, 800 Dongchuan Road, Shanghai, P. R. China

\*Corresponding author: E-mail: wangxing@sjtu.edu.cn

**Abstract** Editorial board members, who are considered the gatekeepers of scientific journals, play an important role in academia, and may directly or indirectly affect the scientific output of a university. In this article, we used the quantile regression method among a sample of 1,387 university in chemistry to characterize the correlation between the number of editorial board members and the scientific output of their universities. Furthermore, we used time-series data and the Granger causality test to explore the causal relationship between the number of editorial board members and the number of articles of some top universities. Our results suggest that the number of editorial board members is positively and significantly related to the scientific output (as measured by the number of articles, total number of citations, citations per paper, and h index) of their universities. However, the Granger causality test results suggest that the causal relationship between the number of editorial board members and the number of articles of some top universities is not obvious. Combining these findings with the results of qualitative interviews with editorial board members, we discuss the causal relationship between the number of editorial board members and the scientific output of their universities.

# 1. Introduction

In academia, the editorial boards of scholarly journals have an important influence on the quality and relevance of published research (Garcia-Carpintero et al. 2010); 'it is considered that the critical mentality and decisions of editorial boards protect the social and intellectual integrity of science' (Braun and Diospatonyi 2005a: 1548). Editorial boards are important to the entire academic world, and there seems to be a possible relationship between a university's representation on a journal's editorial board and its scientific output in that journal. Editorial board members are considered gatekeepers of journals (Braun and Diospatonyi 2005b, 2005a, 2005c; Zsindely et al. 1982). 'These gatekeepers, in controlling the system of manuscript evaluation and selection, occupy powerful strategic positions in the collective activity of the field' (Zsindely et al. 1982: 58). Therefore, they may have an important influence on the scientific output of a university by holding academic influence (i.e., by deciding whether to accept or reject a manuscript, controlling the research themes of a journal, determining the hotspot domain and topics of the subject, and setting the academic evaluation criteria). Editorial board members are also usually eminent researchers; they are normally appointed on the basis of their own strong publication and citation records, and they may directly produce a substantial amount of high-impact scientific output for their affiliated universities. Thus, editorial board members may directly or indirectly affect the international influence of a university's scientific output and may be important to the development and evolution of universities.

Several studies have examined the correlation between the number of editorial board members and scientific output of universities (Braun et al. 2007a; Burgess and Shaw 2010; Chan and Fok 2003; Chan et al. 2005; Gibbons and Fish 1991; Kaufman 1984; Urbancic 2005). The general purpose of these studies was to use the number of editorial board members as an evaluation indicator to measure the scientific output of universities. Some studies found a positive correlation between university ranking based on the number of editorial board members and scientific output, whereas others did not. These studies have mainly focused on the fields of economics and management and have not yet yielded convergent results. In particular, existing studies have left the following gaps.

First, though editorial board members may be important to universities and there may be some relationship between the number of editorial board members and the scientific output of universities, there has been a lack of study about the causality in this relationship. We know little about the direction of causality; that is, whether the number of editorial board members drives the scientific output or the scientific output drives the number of editorial board members. Previous studies on this relationship either calculated correlation coefficients or showed the overlap between the top performers in two types of rankings. However, correlation analysis, including regression analysis, cannot demonstrate a causal relationship. It may be more interesting to determine whether such a causal relationship exists between the number of editorial board members and the scientific output of universities and, if so, which drives which. Unfortunately, this important issue has been ignored in previous studies.

Second, few studies have characterized the correlation between the number of editorial board members and scientific output of universities. The overlap between top performers in two types of rankings says nothing about the medium and low performers. In contrast, ordinary least squares (OLS) regression coefficients estimate the effects of the number of editorial board members on the mean of the conditional distribution of scientific output. There are differences among universities, and it is possible that the relationship between the number of editorial board members and the scientific output of universities is different among universities in different parts of the conditional scientific output distribution. Therefore, showing the overlap between top performers or using OLS models to draw conclusions may result in an incomplete picture of this relationship.

Third, most studies on the topic have used the number of articles published as an indicator of a university's scientific output and have compared this number to the number of editorial board members. Few studies have used indicators related to the impact of the scientific output, such as the total number of citations, to measure the scientific output of universities. Citations per paper and institutional h index (an institution has index h if h of its Np papers have at least h citations each and the other (Np - h) papers have ≤ h citations) (Prathap 2006) have not been used in the existing literature.

In this study, we examined chemistry subject at a sample of 1,387 universities and used the quantile regression method to investigate the variation in the correlation between the number of editorial board members and the scientific output (as measured by number of articles, total number of citations, citations per paper, and h index) across the entire conditional scientific output distribution. Furthermore, we explored the causal relationship between the number of editorial board members and the scientific output of universities. We used time-series data and the Granger causality test to explore the nature of the causal relationship between the number of editorial board members and the number of articles of some top universities. We also interviewed some editorial board members about the causal relationship between the number of editorial board members and the scientific output of universities. We chose to focus on chemistry because the available data about the number of editorial board members in chemistry were relatively complete.

The remainder of this paper is organized as follows. In the next section, we analyze the mechanism behind the relationship between the number of editorial board members and scientific output of universities. Section 3 describes the research method. In section 4, we present the results. First, we conducted an empirical test of the correlation between the number of editorial board members and scientific output. Next, we examined the causal relationship between the number of editorial board members and the number of articles of top universities using the Granger causality test. Then, we interviewed several editorial board members from leading chemical journals on the subject of a causal relationship between the number of editorial board members and the scientific output of universities. Finally, in section 5, we discuss our main conclusions and suggestions for future research.

## 2. Theoretical Foundation and Related Work

As mentioned earlier, there are empirical studies on the correlation between the number of editorial board members and the scientific output of universities (the two variables). However, the presence of statistical correlation cannot be used as an indication of a causal relationship. What is the mechanism of the relationship between the two variables? Could there be a causal relation? If so, which drives which? We supposed that there may be a mutually influencing mechanism between the two variables.

### 2.1 The Influence of Scientific Output on the Number of Editorial Board Members

In theory, editorial board members obtained their positions because of their high research achievements. In other words, only individuals with a strong record of published articles and citations are qualified as candidates for editorial boards. Cole and Cole (1973: 80) have argued that 'unless editorial board members are appointed based on their scientific achievements, the academic community might find it difficult to view their authority as legitimate.' Rynes (2006: 1098) asserted that 'scholars with strong publication and citation records are the most obvious candidates to receive an editorial board invitation to an academic journal.'

Therefore, regarding an individual scholar, their scientific output capacity influences his or her chances of being selected as an editorial board member. Extending this theory to the university level, the greater the quantity and impact of research produced by a university, the greater the probability should be that the university has a higher number of editorial board members. This could be a possible mechanism behind the influence of scientific output on the number of editorial board members at universities.

To our knowledge, there have been few empirical studies on the influence of scientific output of universities on the number of editorial board members. However, many studies have shown that editorial board members had high scientific output records; this has been true in various disciplines, such as medicine (Bakker and Rigter 1985), nanotechnology (Braun et al. 2007b), management (Valle and Schultz 2011), and library and information science (Willett 2013). These studies indirectly indicated that editorial board members were selected based on their outstanding research ability. Valle and Schultz (2011) suggested that it would be unlikely for an editorial board member to get their paper published using their influence as a board member. It was more likely, the researchers argued, that editorial board members had a high number of published articles because of their own research abilities. Although some empirical studies have indicated that the scientific output records of editorial board members were not as high as expected, such as in the field of social work (Pardeck 1992; Pardeck and Meinert 1999), it has become a mainstream assumption in academia that editorial board members tend to have a higher research ability.

## 2.2 The Influence of the Number of Editorial Board Members on Scientific Output

There are two possible reasons for the influence of the number of editorial board members at a university on the university's scientific output.

First, editorial board members may produce a substantial amount of high-impact scientific output for their universities owing to their outstanding research abilities. As mentioned in the above section on the influence of scientific output on the number of editorial board members, scholars are, at least theoretically, selected as editorial board members because they have demonstrated research achievements (e.g., strong publication and citation records). Further, it has been shown that board members usually have a higher research ability than their peers. Therefore, board members could contribute a certain amount of high-impact scientific output to their universities because of their own high research abilities, independently of their status as board members. These are actually two aspects of the same issue.

Second, editorial board members, considered the gatekeepers of scholarly journals, may have a certain influence on the scientific output of their universities by controlling the academic discourse (e.g., controlling the research hotspots of their respective fields and the themes of journal articles, making decisions to publish journal articles, and setting the academic evaluation criteria of journals). Authors with similar academic backgrounds to these editorial board members (e.g., working in the same institution or graduated from the same institution) might share similar academic viewpoints, research topics, or research directions. Further, they might have similar preferences in research methods or paradigms. Owing to this conformity, authors from the same institutions as editorial board members might acquire academic recognition more easily, and, therefore, their articles might be more likely to be published. Moreover, the research topic preferences of editorial boards might influence the research hotspots in the field, and might result in higher numbers of citations of articles related to these topics.

Currently, there is, to our knowledge, no research that directly demonstrates that articles written by scholars from the same institutions of the editorial board members are more likely to be published and to have more citations owing to academic discourse controlled by board members. However, some scholars have investigated similar topics from indirect angles. For examples, Brogaard et al. (2014) found that, in the field of economics and finance, the articles written by colleagues of the editorial board members tended to have a higher acceptance rate, as well as a higher number of citations. Other empirical research has shown that articles written by authors from the same institutional background as editorial board members tend to have more citations (Laband and Piette 1994; Medoff 2003).

## 3. Methodology

### 3.1 Samples for Correlation Analysis

The source for the journal sample used in our correlation study was the category of chemistry from Thomson Reuters's 2011 Journal Citation Reports (Science Edition), comprising 514 journals that serve seven sub-disciplines: chemistry science, analytical; chemistry science, applied; chemistry science, inorganic & nuclear; chemistry science, medicinal; chemistry science, multidisciplinary; chemistry science, organic; and chemistry science, physical. Between April and July, 2013, we visited the homepage of each journal to manually collect the names and affiliations of the gatekeepers having positions such as editor-in-chief, chief editor, co-editor, deputy editor, associate editor, regional editor, senior editor, editorial board member, or advisory board member. Furthermore, following Gibbons and Fish (1991), if editorial board members were affiliated with multiple institutions, we counted their names as many times as they appeared. From the total number of chemistry journals, 118 did not specify the names or affiliations of their editorial board members. Eventually, we acquired the information of 10,121 board members at 396 journals, who were affiliated with 1,387 universities. These universities constituted the sample for the correlation analysis of the number of editorial board members and the scientific output.

To acquire data about the scientific output of the 1,387 universities, during October 2013 we used the advanced search section of the Thomson Reuters Web of Science. For example, to obtain the scientific output of Stanford University, we used the following search format: WC = (CHEMISTRY ANALYTICAL OR CHEMISTRY APPLIED OR CHEMISTRY INORGANIC NUCLEAR OR CHEMISTRY MEDICINAL OR CHEMISTRY MULTIDISCIPLINARY OR CHEMISTRY ORGANIC OR CHEMISTRY PHYSICAL) AND OG = (Stanford University). 'WC' and 'OG' refer to the discipline and name of the organization in Web of Science, respectively. In addition, we limited results to scientific output published between 2008 and 2012. Subsequently, we obtained data about the number of articles, total number of citations (citations from 2008 to 2012 for articles published from 2008 to 2012), citations per paper (same as above), and h index (same as above) for the 1,387 universities by creating citation reports.

The citation windows of five years and less are used here. It should be noted that the Spearman correlation between the long term (31 year) citation counts and the cumulative citation counts in years 1 to 5 are 0.266, 0.592, 0.754, 0.830, and 0.872, respectively; moreover, the potential error of using a short time window will be higher for highly cited papers (Wang 2013). However, Wang's (2013) study was conducted at an individual paper level, not at an institutional level. 'If we can assume that the shares of slow and fast ageing articles are the same for all focal authors and institutions to be evaluated, then using short citation time windows would penalize every evaluatee equally and therefore is less problematic for evaluation purposes' (Wang 2013: 867). Some studies also suggested that using short citation time windows did not change the evaluation results much at the institutional level (Glanzel 2008; Abramo et al. 2012). Therefore, the potential errors of using a short time window may not play a major role in our study.

Moreover, with the purpose of reducing the fluctuations in citations per paper from universities with a small number of publications and obtaining a statistically valid and reliable analysis, we used an arbitrary threshold of 441 articles (which is 10% of the mean value of the total number of publications produced by the five most productive universities) for calculating citations per paper. In total, we analyzed 531 universities for the correlations between the number of editorial board members and the citations per paper.

## 3.2 Samples for Granger Causality Test

We also collected time-series data for the Granger causality test using two variables: the number of editorial board members and the number of articles published per university. Since it was difficult to acquire time-series data for the number of editorial board members of all 1,387 universities for all journals, we referred to studies conducted by Brown (2007) and selected the following nine top journals as sample journals for this analysis: *Journal of the American Chemical Society*, *Angewandte Chemie International Edition, Chemical Reviews*, *Accounts of Chemical Research*, *Analytical Chemistry*, *Biochemistry*, *Chemistry of Materials*, *Inorganic Chemistry*, and *Journal of Organic Chemistry*. Since the majority of these journals did not reveal the affiliations of their editorial board members until 1998, we chose the period from 1998 to 2014.

We faced the risk that there might be a limited number of board members from each university every year at these nine journals, so we used the Shanghai ranking's top 20 universities (their number of editorial board members was higher every year) in chemistry as our sample universities for the Granger causality test. First, we recorded and calculated the number of editorial board members over the years of 20 universities from 1998 to 2014 at the nine journals. Next, to minimize statistical errors caused by a limited number of board members from each university every year and to obtain more reliable results, we artificially set a threshold: only universities with no less than five editorial board members for at least one year were considered qualified for our Granger causality test. Fourteen universities constituted the final sample for this analysis.

Using the aforementioned advanced search function of Thomson Reuters Web of Science, we obtained the number of articles published in the nine journals each year from 1998 to 2014 at 14 universities. Data for both the number of editorial board members and the number of articles were collected in June 2015.

### 3.3 Samples for E-mail Interview

To deepen our understanding of the relationship between the number of editorial board members and the scientific output, we conducted semi-structured interviews with several board members from the nine journals. Considering that most of the board members resided outside of China, interviews were conducted via e-mail. The semi-structured interview guide included questions regarding possible causal mechanisms (aforementioned; see the theoretical foundation section) between the number of editorial board members and the scientific output of universities. There were five questions included (Appendix A).

Each editor-in-chief and the deputy editor of the nine journals were the potential subjects of our interviews. In addition, since there were many other editorial board members, we used the website www.random.org/nform.html as a tool to randomly select three other members of each journal as potential interview subjects. Next, we used Google to reach the official websites of the universities with which these board members were affiliated, from which websites we sought to obtain the board members' e-mail addresses. Next, we sent individual e-mails containing the interview questions to each of the selected subjects. In total, 130 e-mails were sent. We set one month as our response period, and 16 board members answered our interview questions within this time range. All e-mails were sent in July 2015. The number of respondents per journal is presented in Table 1.

**Table 1.** Number of respondents per journal.

| Journal Name | Number of respondents | Number of chief editor or deputy editor respondents |
|---|---|---|
| *Accounts of Chemical Research* | 0 | 0 |
| *Analytical Chemistry* | 5 | 4 |
| *Angewandte Chemie International Edition* | 3 | 2 |
| *Biochemistry* | 1 | 1 |
| *Chemical Reviews* | 2 | 2 |
| *Chemistry of Materials* | 0 | 0 |
| *Inorganic Chemistry* | 0 | 0 |
| *Journal of Organic Chemistry* | 2 | 2 |
| *Journal of the American Chemical Society* | 3 | 2 |
| Total | 16 | 13 |

### 3.4 Granger Causality Test Models

In this study, we used the Granger causality test to investigate the causal relationship between the number of editorial board members and the number of articles published at top universities. The Granger causality test is a statistical hypothesis model for causal prediction based on time-series data (Granger 1969). It is a commonly used method for examining causal relationships between variables in economics and management.

The basic principle of the Granger causality test is as follows: To examine whether a variable X is the cause of another variable Y, a restricted regression model such as formula (1) should be established first to show that Y can be explained by its own past values. Then, past values of X as the explanatory variable are introduced into the formula (1) to form an unrestricted regression model, yielding formula (2). If introducing past values of X can significantly improve the prediction level of Y, then X is said to be the Granger cause of Y. Similarly, these steps can be repeated to determine whether Y causes X.

$$Y_t = \alpha_0 + \sum_{i=1}^{m} \alpha_i Y_{t-i} + \mu_t \tag{1}$$

$$Y_t = \alpha_0 + \sum_{i=1}^{m} \alpha_i Y_{t-i} + \sum_{j=1}^{m} \beta_j X_{t-j} + \mu_t \tag{2}$$

In our study, X represents the number of editorial board members and Y represents the number of articles. $\alpha_0$ is a constant, $\mu_t$ is a white noise sequence, $\alpha_i$ and $\beta_j$ are coefficients, and m is the number of lagged terms. For both formulas (1) and (2), the longer the lag length, the better it reveals the dynamic features of the models. However, if the lag length is too long, the freedom of the model will be reduced. There needs to be a balance between the two variables. Moreover, from the perspective of actual publishing cycles, the publishing cycle of articles from the American Chemical Society is 4–8 months; from the perspective of editorial board members as a research manpower input, some scholars choose a lag of 1–3 years (Brogaard et al. 2014; Yu 2013). However, from the perspective of the period when the editorial board members obtained their positions, there is no fixed standard. Based on the above factors and for the sake of prudence, we selected a lag length of 1–4 years for our test.

To date, the Granger causality test model has been applied to a number of studies in the field of scientometrics (Inglesi-Lotz et al. 2014; Lee et al. 2011). Eviews 6.0 software was used for all statistical analyses.

## 4. Results and Discussion

### 4.1 Analysis of Ordinary Least Squares Regression and Quantile Regression

Table 2 contains the descriptive statistics of the dependent and independent variables selected for this study. In our study, the independent variable was measured as the number of editorial board members affiliated to a particular university; the dependent variables were measured by the number of articles, total number of citations, citations per paper, and h index.

**Table 2.** Descriptive statistics of dependent/independent variables.

| Variable | Mean | SD | Median | Minimum | Maximum |
|---|---|---|---|---|---|
| Panel A: Dependent variable | | | | | |
| Number of articles | 505.27 | 631.65 | 283.00 | 0.00 | 4870.00 |
| Total number of citations | 4938.69 | 7502.10 | 2125.00 | 0.00 | 63009.00 |
| Citations per paper | 9.76 | 3.45 | 9.52 | 2.99 | 24.68 |
| H index | 23.44 | 14.77 | 21.00 | 0.00 | 94.00 |
| Panel B: Independent variable | | | | | |
| Editorial board members | 7.30 | 11.45 | 3.00 | 1.00 | 118.00 |

First, we estimated four OLS models, for which the results are shown in Table 3. As expected, the results show that the number of editorial board members is positively and significantly ($p < 0.01$) related to the scientific output (as measured by the number of articles, total number of citations, citations per paper, and h index) from their universities.

**Table 3.** Results of the OLS regressions.

| | Coeff. | SE | $t$ statistic | $P$ value |
|---|---|---|---|---|
| Panel A: Dependent variable (number of articles) | | | | |
| Editorial board members | 39.072[*] | 1.047 | 37.318 | 0.000 |
| Constant | 220.159[*] | 14.210 | 15.494 | 0.000 |
| $R^2$ | | | | 0.501 |
| $F$-test | | | $F = 1392.597$ ($P = 0.000$) | |
| Breusch-Pagan-Godfrey test | | | $\chi^2 = 64.341$ ($P = 0.000$) | |
| Panel B: Dependent variable (total number of citations) | | | | |
| Editorial board members | 532.534[*] | 10.265 | 51.878 | 0.000 |
| Constant | 1052.762[*] | 139.311 | 7.557 | 0.000 |
| $R^2$ | | | | 0.660 |
| $F$-test | | | $F = 2691.370$ ($P = 0.000$) | |
| Breusch-Pagan-Godfrey test | | | $\chi^2 = 141.252$ ($P = 0.000$) | |
| Panel C: Dependent variable (citations per paper) | | | | |
| Editorial board members | 0.123[*] | 0.008 | 15.080 | 0.000 |
| Constant | 7.906[*] | 0.175 | 45.058 | 0.000 |
| $R^2$ | | | | 0.301 |
| $F$-test | | | $F = 227.421$ ($P = 0.000$) | |
| Breusch-Pagan-Godfrey test | | | $\chi^2 = 21.231$ ($P = 0.000$) | |
| Panel D: Dependent variable (h index) | | | | |
| Editorial board members | 0.957* | 0.023 | 41.121 | 0.000 |
| Constant | 16.458[*] | 0.316 | 52.107 | 0.000 |
| $R^2$ | | | | 0.550 |
| $F$-test | | | $F = 1690.913$ ($P = 0.000$) | |
| Breusch-Pagan-Godfrey test | | | $\chi^2 = 128.839$ ($P = 0.000$) | |

*Note.* * significant at 1% significance level.

However, the Breusch-Pagan-Godfrey test demonstrated the presence of heteroscedasticity in the four OLS regression models (Table 3). Heteroscedasticity was also revealed by the scatter plot of the number of editorial board members and the scientific output indices. For example, Figure 1 shows that the dispersion of the number of articles increased with the number of editorial board members, which is a typical characteristic of heteroscedasticity. If we trace these divergent trajectories, we might get regression lines with different slopes depending on whether we consider the higher quantile (i.e., dotted line R1) or lower quantile (dotted line R2) of the conditional scientific output distribution, and R is the regression line calculated based on the results of OLS. In other words, the relationship between the number of editorial board members and the scientific output of universities may be different when the universities are in different quantiles of the conditional scientific output distribution. Considering this feature of our data, to gain a better understanding and obtain a more complete picture of the relationship between the number of editorial board members and the scientific output of universities, we conducted a quantile regression analysis.

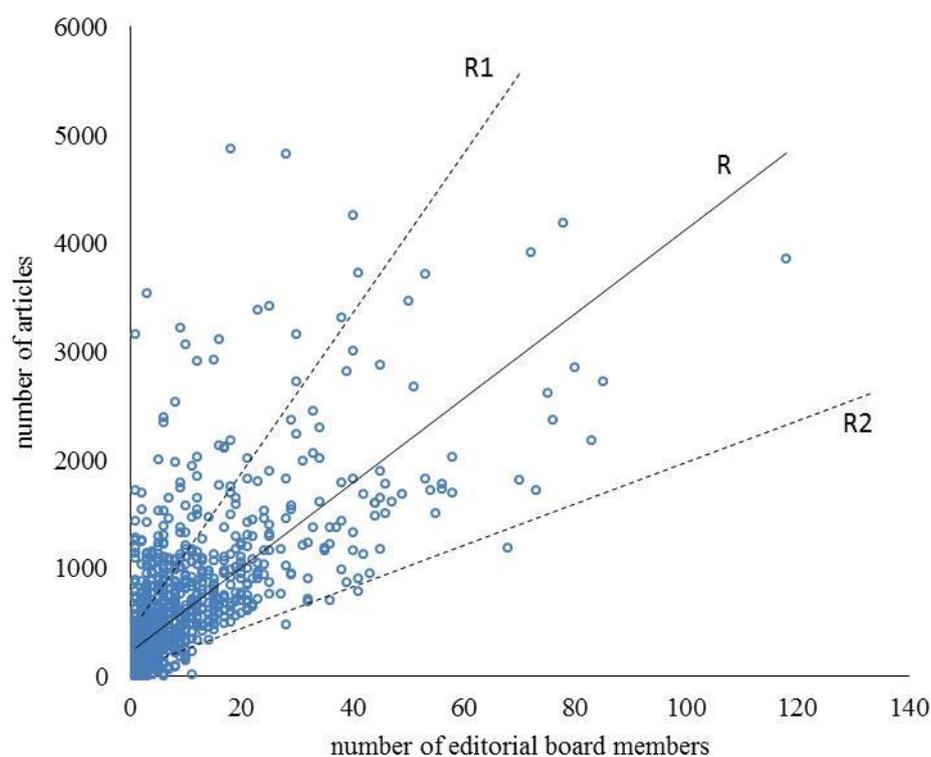

**Figure 1.** The heteroskedastic relationship between the number of editorial board members and the number of articles.

We computed 19 quantile regression estimates (5%−95%, in 5% increments). Figures 2−5 plot the coefficient values at various quantiles of the number of articles, total number of citations, citations per paper, and h index, respectively, together with the OLS estimates. Specific data for each quantile can be found in Appendix B.

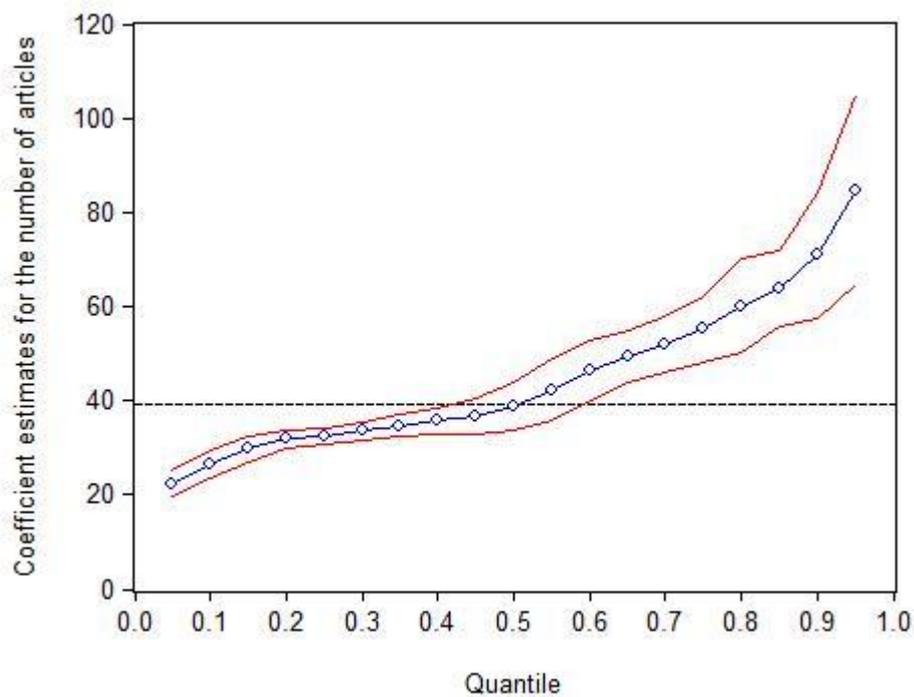

**Figure 2.** Coefficient estimates for the number of articles, with 95% confidence intervals. The horizontal axis represents the quantile; the vertical axis shows the estimated parameter. The dashed horizontal line represents the OLS estimate. The estimated parameters of the quantile regression are displayed as circles. The 95% confidence intervals are displayed as solid lines.

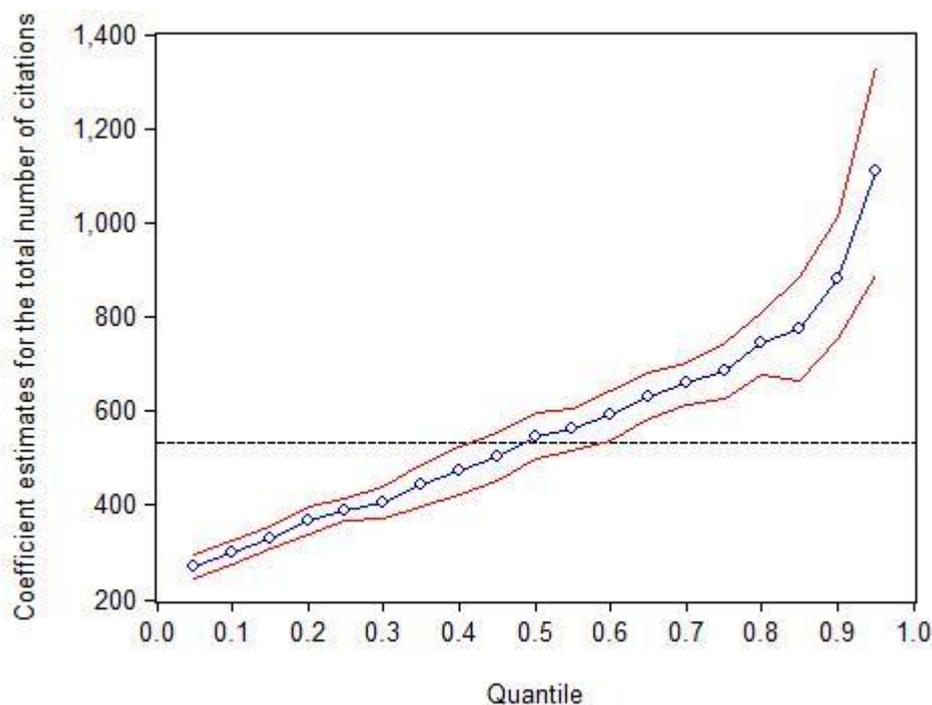

**Figure 3.** Coefficient estimates for the total number of citations, with 95% confidence intervals. The horizontal axis represents the quantile; the vertical axis shows the estimated parameter. The dashed horizontal line represents the OLS estimate. The estimated parameters of the quantile regression are displayed as circles. The 95% confidence intervals are displayed as solid lines.

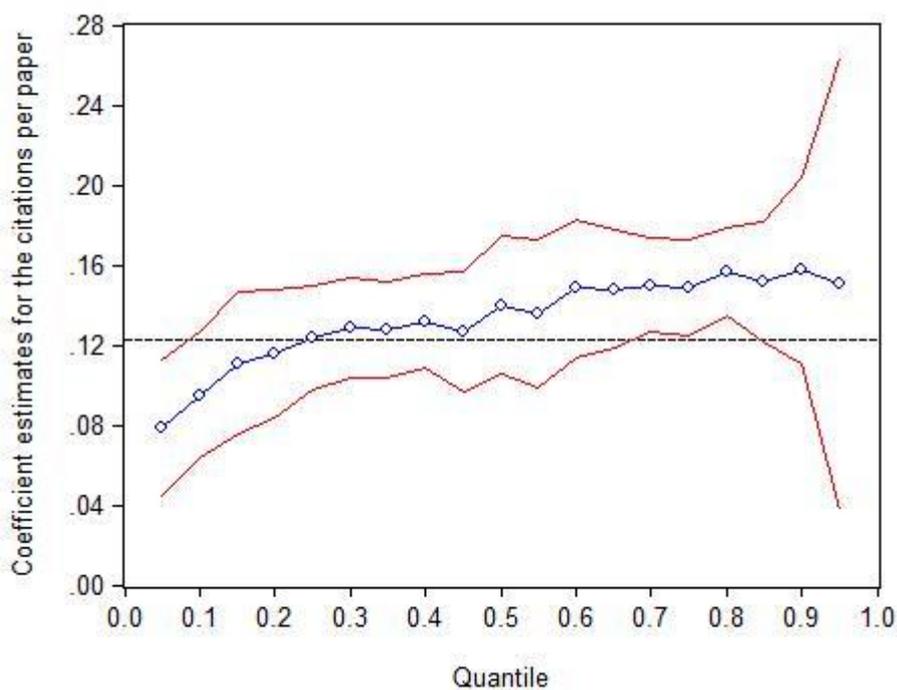

**Figure 4.** Coefficient estimates for the citations per paper, with 95% confidence intervals. The horizontal axis represents the quantile; the vertical axis shows the estimated parameter. The dashed horizontal line represents the OLS estimate. The estimated parameters of the quantile regression are displayed as circles. The 95% confidence intervals are displayed as solid lines.

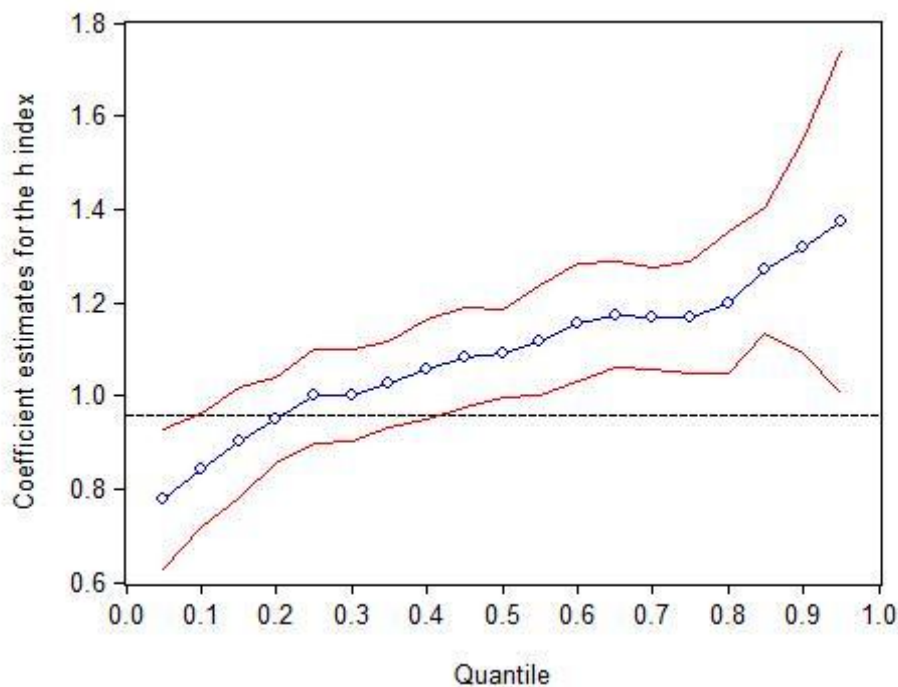

**Figure 5.** Coefficient estimates for the h index, with 95% confidence intervals. The horizontal axis represents the quantile; the vertical axis shows the estimated parameter. The dashed horizontal line represents the OLS estimate. The estimated parameters of the quantile regression are displayed as circles. The 95% confidence intervals are displayed as solid lines.

Quantile regression results show that the coefficient for editorial board members in relation to the scientific output (any of the number of articles, total number of citations, citations per paper, or h index) is positive and significant at the 1% level for all the quantiles. Furthermore, the coefficient gradually increases from the lower to the upper quantiles of the conditional distribution of scientific output (any of the number of articles, total number of citations, citations per paper, or h index), which indicates that the coefficient of the number of editorial board members on the indicators of scientific output is greater (smaller) when the university is at the higher (lower) quantile of the conditional distribution of scientific output. We should note here that the editorial board coefficient of the quantile regression that relates to the number of articles and total number of citations is above the OLS results at the higher quantiles, and below the OLS results at the lower quantiles. The OLS models underestimate the editorial board members coefficient that relates to the citations per paper and h index for the 25% quantile and above. Therefore, quantile regression models reveal more information than is evident in the OLS models.

The interpretation of the results from quantile regressions will be discussed in combination with the results of the Granger causality test in the next section.

## 4.2 Analysis of Granger Causality Test

The prerequisite of the Granger causality test is that the two series are stationary or co-integrated, otherwise the problem of 'spurious regression' might occur. Therefore, it was necessary to conduct unit root and co-integration tests in the number of editorial board members and number of articles time series for the 14 universities. For this purpose, augmented Dickey-Fuller (ADF) and Johansen co-integration tests were used. Eventually, we conducted the Granger causality test on the nine universities whose number of editorial board members and number of articles either satisfied the condition of being stationary or being co-integrated (Table 4).

**Table 4.** Granger cause test between the number of editorial board members and the number of articles of universities (1998-2014).

| Rank | University | EB | PUB | Co-integration | EB→PUB | PUB→EB | Lag |
|------|-----------|-----|------|------|------|------|-----|
| 1 | UC-Berkeley | I(1) | I(1) | √ | 1.669 （0.316） | 1.975 （0.263） | 4 |
| 2 | Harvard University | I(1) | I(0) | | / | / | / |
| 3 | Stanford University | I(0) | I(0) | | 1.697 （0.215） | 1.952 （0.186） | 1 |
| 4 | Northwestern University | I(2) | I(1) | | / | / | / |
| 6 | MIT | I(1) | I(1) | √ | 1.269 （0.323） | 2.098 （0.174） | 2 |
| 7 | California Institute of Technology | I(0) | I(1) | | / | / | / |
| 8 | ETH-Zurich | I(0) | I(0) | | 0.704 （0.518） | 5.162$^{**}$ （0.029） | 2 |
| | | | | | 0.693 （0.585） | 4.443$^{**}$ （0.048） | 3 |
| | | | | | 0.507 （0.737） | 5.283$^{*}$ （0.068） | 4 |
| 9 | Kyoto University | I(0) | I(0) | | 1.384 （0.380） | 1.246 （0.418） | 4 |
| 10 | UCLA | I(1) | I(1) | √ | 5.644$^{**}$ （0.034） | 1.529 （0.238） | 1 |
| | | | | | 4.693$^{**}$ （0.037） | 2.211 （0.160） | 2 |
| 11 | University of Pennsylvania | I(0) | I(0) | | 0.812 （0.578） | 49.647$^{***}$ （0.001） | 4 |
| 12 | Yale University | I(0) | I(0) | | 0.001 （0.971） | 0.621 （0.445） | 1 |
| 13 | UC-Santa Barbara | I(1) | I(0) | | / | / | / |
| 18 | UC-San Diego | I(1) | I(1) | √ | 2.017 （0.257） | 0.282 （0.876） | 4 |
| 20 | University of Tokyo | I(0) | I(1) | | / | / | / |

*Note*. EB, PUB represent respectively the number of editorial board members and the number of articles.

Column 1 is the 2014 Shanghai ranking of the selected universities in chemistry.

Columns 3 and 4 state, respectively, the data characteristics of the EB and Pub serials. If the series itself is stationary, we represent it by I (0). If the series are integrated with order n, we represent it by I (n).

Column 5 uses '√' to indicate that the two series were co-integrated.

Columns 6 and 7 represent the *F*-value of the Granger causality test. *P*-value in parentheses: '***', '**', and '*' significant at 1%, 5%, and 10% significant level.

Column 8 is the lag phase. For the three universities that had a significant causal relationship between the two variables, the lag phases having a significant causal relationship were all provided. For universities with no detected causal relations during lag phase 1 to 4, we only present an optimal lag phase based on the Akaike Information Criterion.

The results showed that for the University of California, Los Angeles (UCLA), the number of editorial board members was the Granger cause of the number of articles in lag phases 1 and 2. By contrast, for Swiss Federal Institute of Technology Zurich (ETH-Zurich), the number of articles was the Granger cause for the number of editorial board members in lag phases 2, 3, and 4. For the University of Pennsylvania, the number of articles was the Granger cause of the number of editorial board members in lag phase 4. There was no significant causal relationship in either direction for the other six universities. The regression equations of UCLA, ETH-Zurich, and the University of Pennsylvania based on the Granger causality test model (2) are shown in Table 5.

**Table 5.** Regression equation based on Granger causality test model of the three universities where Causal relationships were detected.

| Lag | Regression equation based on Granger causality test model |
|---|---|
| | Part A: UCLA |
| 1 | $Y_t = \mathbf{\textit{91.979}} + 0.379\ Y_{t-1} - \mathbf{\textit{4.508}}\ X_{t-1}$ |
| 2 | $Y_t = \mathbf{\textit{152.597}} + 0.150\ Y_{t-1} - 0.078\ Y_{t-2} - \mathbf{\textit{4.405}}\ X_{t-1} - 4.053\ X_{t-2}$ |
| | Part B: ETH-Zurich |
| 2 | $X_t = \mathbf{\textit{4.700}} + \mathbf{\textit{0.622}}\ X_{t-1} - 0.374\ X_{t-2} + \mathbf{\textit{0.054}}\ Y_{t-1} - \mathbf{\textit{0.062}}\ Y_{t-2}$ |
| 3 | $X_t = 4.583 + \mathbf{\textit{0.902}}\ X_{t-1} - 0.529\ X_{t-2} - 0.168\ X_{t-3} + \mathbf{\textit{0.050}}\ Y_{t-1} - \mathbf{\textit{0.085}}\ Y_{t-2} + 0.033\ Y_{t-3}$ |
| 4 | $X_t = 8.617 + 0.533\ X_{t-1} - 0.185\ X_{t-2} - 0.334\ X_{t-3} - 0.501\ X_{t-4} + \mathbf{\textit{0.039}}\ Y_{t-1} - \mathbf{\textit{0.060}}\ Y_{t-2} - 0.005\ Y_{t-3} + 0.019\ Y_{t-4}$ |
| | Part C: University of Pennsylvania |
| 4 | $X_t = \mathbf{\textit{5.617}} + 0.304\ X_{t-1} + \mathbf{\textit{0.625}}\ X_{t-2} - 0.069\ X_{t-3} + \mathbf{\textit{0.777}}\ X_{t-4} - 0.007\ Y_{t-1} - \mathbf{\textit{0.025}}\ Y_{t-2} - \mathbf{\textit{0.027}}\ Y_{t-3} - \mathbf{\textit{0.042}}\ Y_{t-4}$ |

*Note.* X, Y represents respectively the number of editorial board members and the number of articles. Bold and italics denote the associated *p*-values lower than 5%.

Although the Granger causality test results suggested unidirectional causality, either from the number of editorial board members to the number of articles or the reverse, in the above three universities, the established regression equations of these three universities based on the Granger causality test model (2) were contradictory with respect to meaning. For example, the coefficient of $X_{t-1}$ was significantly negative ($p < 0.05$) in the equation of lag phase 1 of UCLA, and the coefficient of $X_{t-1}$ was also significantly negative ($p < 0.05$) in the equation of the lag phase 2 of UCLA, indicating that when the number of editorial board members from UCLA increased in the previous phase, the number of articles published would decrease in the current phase. This indication was apparently contradictory. Similar issues occurred in the situation of ETH-Zurich and the University of Pennsylvania.

Based on the above results, there was no apparent causal relationship between the number of editorial board members and the number of articles from the universities we selected for the Granger causality test, which was different from our prior hypothesis of causality. The results differing from the hypothesis may be attributed to the following two reasons.

First, the annual changes in the numbers of editorial board members in the tested universities were not obvious. For example, the minimum number of editorial board members per year from Stanford University was five, while the maximum number was nine. During the period of 1998–2014, the number of editorial board members changed only insignificantly, and it was therefore not easy to show a good corresponding relation with the number of articles, where there were more obvious changes. As a result, it was difficult to detect a causal relation between the two variables.

Second, the causal relationship between the number of editorial board members and the number of articles of universities might not be 'rigid'. In other words, an increase or decrease of one variable does not necessarily cause a significant increase or decrease of the other. The universities we selected were the world's leaders in chemistry, and increasing or reducing one or two editorial board members at these universities might have little effect on the number of articles of these universities. The changes in both the number of editorial board members and the number of articles could be a result of the combined influences of several factors, such that the number of editorial board members or the number of articles was only one of several factors.

Research funding, research personnel input, and research policy could also be factors that affect a university's scientific output. Therefore, although the number of editorial board members could be the same, the influence of this variable on universities' scientific output could differ, (e.g., the regression lines with different slopes shown in Figure 1). This indicates the existence of other possible influencing factors, such as research input or research policy. The difference between universities in research input or research policies may be the cause of the difference in the relation between the number of editorial board members and scientific output at different conditional scientific output distribution quantiles. Since it is difficult to acquire the research input data and research policies of universities in a single discipline, we did not include these factors in our model.

From the perspective of factors affecting the selection of board members, although board members were usually excellent scholars with outstanding research ability, there were still other factors influencing their selection. For example, since editorial board members had to review many manuscripts, which might take time that could be used to conduct scientific research, some excellent scientists might choose not to serve as editorial board members. In addition, geographical factors and reviewer experience were also factors considered in board member selection.

## 4.3 Analysis of Interviews with Editorial Board Members

The following sections summarize the responses of the editorial board members to the interview questions, which included:

- Do editorial board members have an influence on academic discourses?

- Is there any misuse of editorial power?

- Do editorial board members have strong publication and citation records, and are these criteria for selecting board members?

- Is there a causal relationship between the number of editorial board members and the scientific output of universities?

All content inside quotation marks is directly quoted from respondents' interviews.

### 4.3.1 Do Editorial Board Members have an Influence on Academic Discourses?

Most respondents believed that editorial board members had no or only a little influence on academic discourses. They thought that the influence of academic discourses on the scientific output of a university should not be overstated. This influence may be limited, with the publication of an article instead depending on whether there are new discoveries and contributions in the article, as well as the quality of articles themselves. As mentioned in the results and discussion of the Granger causality test, editorial board members' academic discourses might be only one of several factors that influence a university's scientific output; other factors, such as research funding, human input, and research policy, may also play important roles.

However, several respondents pointed out that board members had an influence on the themes or research fields of the articles selected for the journal. For example:

'Editorial board members do have influence on the choice of themes for the journal; that is one of their primary tasks at the Editorial Board meetings'; 'Each journal needs to define which direction of research is relevant for the journal and which direction one thinks will be important in the future. Such things are always subjective, and it is clear that as the editors define the direction of the journal, they have an influence about which type of research is published.'

*4.3.2 Is there any Misuse of Editorial Power?* It should be noted that although board members may influence academic discourses, this does not mean that they misuse their own journals to help themselves or their universities publish unworthy articles. Our emphasis was on factors such as controlling the themes and research direction of journals, setting the academic evaluation criteria of journals, and preference in certain academic viewpoints or research paradigms.

The majority of the respondents believed that there was little misuse of power among editorial board members, or that this phenomenon was rare in their journals, which confirms the results of previous studies (Bosnjak et al. 2010; Frandsen and Nicolaisen 2011; Sugimoto and Cronin 2012). Combining the answers given by the respondents, we found three reasons that respondents thought it unlikely that board members of SCI journals abuse their editorial power. First, journals' peer review rules prevent an editorial board member from processing or reviewing their own articles or any articles that might have a conflict of interest, including articles from their colleagues or students. Second, since editorial board members usually have a better record of publication and citations and are capable of producing high-quality articles, abusing their power to publish their own papers would be unnecessary, and even harmful to their academic reputations. Third, journal articles will eventually be published and supervised by peers. Scientific journals have been around for approximately 350 years, and it is almost impossible to imagine that they could have survived for so long if articles were mainly published for egoistical reasons. If this were the case, they would not be read by scientists (Frandsen and Nicolaisen 2011).

### 4.3.3 Do Editorial Board Members have Strong Publication and Citation Records, and are these Criteria for Selecting Board Members?

Nearly all respondents believed that the editorial board members of their journals had strong publication and citation records. This result was in line with the mainstream belief in the academic community and the findings of some previous research (Bakker and Rigter 1985; Braun et al. 2007b; Valle and Schultz 2011; Willett 2013).

As for the selection criteria of editorial board members, having a strong publication and citation record was the most frequently mentioned (10 respondents). Therefore, we considered that having a strong publication and citation record is a basic standard or prerequisite in the selection of editorial board members. Other factors may also affect board member selection. The selection of the editorial board is a comprehensive process, taking various factors into consideration. In addition to having a strong publication and citation record, respondents mentioned the following criteria: academic prestige (seven respondents), research fields (three respondents), geographical location (three respondents), experience as a reviewer (two respondents), and contributions to the journal (one respondent).

### 4.3.4 Is there a Causal Relationship between the Number of Editorial Board Members and the Scientific Output of Universities?   Most respondents believed that there was no causal relationship between the number of editorial board members and the scientific output of universities, or that there was a non-causal correlation between the two variables. Thus the interview results confirmed the results of the Granger causality test and the regression analysis. As discussed in the results of the Granger causality test, the causal relationship between the number of editorial board members and the scientific output of universities might not be rigid, i.e. an increase or decrease of one variable does not necessarily cause a significant increase or decrease of the other. Furthermore, the positive and significant correlation between the number of editorial board members and the scientific output of universities revealed by the regression results may reflect some respondents' opinion that 'strong universities tend to have strong researchers who are more likely to be recruited as editors of international journals.'

Since editorial board members, as the gatekeepers of journals, differ from other types of research personnel, most respondents did not note a causal relationship between the two variables because they were considering the causal relationship from the perspective that editorial board members control the academic discourse. From this perspective, the causal relationship between the two variables may be unobvious, confirmed by most respondents' responses to the first interview question. Therefore, from the perspective that editorial board members control the academic discourse, we believe that board members' academic discourse might influence a university's scientific output, but to a limited extent.

Moreover, we believe that editorial board members usually have high research ability and are mainly selected based on their strong publication and citation records, which is also confirmed by the responses to the third interview question. From this we further speculate that the greater the quantity and impact of research produced by a university, the more chance there is that the university has a higher number of editorial board members. Similarly, board members could also contribute a certain amount of high-impact scientific output to the university because of their own high research ability. Therefore, we speculate that there may be mutual causality between the number of editorial board members and the scientific output of universities from the perspective that editorial board members have high research ability.

## 5. Conclusion

In this study, we examined the relationship between the number of editorial board members and the scientific output of universities in the field of chemistry. For the study period (2008–2012), our empirical results indicated that the number of editorial board members is positively and significantly related to the quantity and impact of the scientific output (as measured by the number of articles, total number of citations, citations per paper, and h index) of their universities. In addition, we observed from quantile regressions that the coefficient of the number of editorial board members on the indicators of scientific output is greater (smaller) when the university is at a higher (lower) quantile of the conditional distribution of scientific output. Our quantile regressions revealed information that was not observable with the OLS model.

More importantly, our Granger causality test results show that the causal relationship between the number of editorial board members and the number of articles of some top universities in chemistry is not obvious. Furthermore, the interview results regarding the causal relationship between the number of editorial board members and the scientific output of universities confirmed the results of the Granger causality test and the regression analysis.

There are some limitations to note, which also suggest directions for future research. First, this study used the total number of citations, citations per paper, and h index to measure the impact of scientific output of universities. However, it should be noted that citation as an evaluation indicator is not a direct reflection of the quality or importance, or even the impact, of scientific output; it only partially reflects the impact of scientific output because it is determined partly by the impact of the paper on the advance of scientific knowledge, but also partly by other factors, including various social and political pressures (Martin and Irvine 1983). Future studies could investigate other evaluation indicators, such as h-like index or peer evaluation, to obtain convergent results and diminish the impact of other factors. Second, some details of the research await improvement. For example, the editorial board members could be divided into three groups: the editor-in-chief, the deputy editor, and other editorial board members. Additionally, different journals have different levels of academic influence, among which we did not distinguish in this study. Moreover, because of the difficulties of data collection, we did not include other variables influencing scientific output in our regression analysis. Future studies could further categorize the board members according to specific positions and journals according to their academic influence, as well as introduce more variables related to scientific output, to reach a more accurate result. Third and finally, our empirical results are limited to chemistry. More studies are needed to investigate whether our conclusions can be extended to other disciplines.

## References

Abramo, G., Cicero, T. and D'Angelo, C. A. (2012) 'A Sensitivity Analysis of Research Institutions' Productivity Rankings to the Time of Citation Observation', *Journal of Informetrics*, 6/2, 298–306.

Bakker, P. and Rigter, H. (1985) 'Editors of Medical Journals: Who and from where', *Scientometrics,* 7, 11–22.

Bosnjak, L. *et al*. (2011) 'Analysis of a Number and Type of Publications that Editors Publish in their own Journals: Case Study of Scholarly Journals in Croatia', *Scientometrics*, 86/1, 227–33.

Braun, T. and Diospatonyi, I. (2005a) 'Counting the Gatekeepers of International Science Journals a Worthwhile Science Indicator', *Current Science*, 89/9, 1548–51.

——. (2005b) 'The Counting of Core Journal Gatekeepers as Science Indicators

really Counts. The Scientific Scope of Action and Strength of Nations', *Scientometrics*, 62/3, 297–319.

——. (2005c) 'Gatekeeping Indicators Exemplified by the Main Players in the International Gatekeeping Orchestration of Analytical Chemistry Journals', *Journal of the American Society for Information Science and Technology*, 56/8, 854–60.

Braun, T. *et al.* (2007a) 'Journal Gatekeepers Indicator-Based Top Universities of the World, of Europe and of 29 Countries—A Pilot Study', *Scientometrics*, 71/2, 155–78.

——. (2007b) 'Gatekeeping Patterns in Nano-Titled Journals', *Scientometrics*, 70/3, 651–67.

Brogaard, J., Engelberg, J. and Parsons, C. A. (2014) 'Networks and Productivity: Causal Evidence from Editor Rotations', *Journal of Financial Economics*, 111/1, 251–70.

Brown, C. (2007) 'The Role of Web-Based Information in the Scholarly Communication of Chemists: Citation and Content Analyses of American Chemical Society Journals', *Journal of the American Society for Information Science and Technology*, 58/13, 2055–65.

Burgess, T. F. and Shaw, N. E. (2010) 'Editorial Board Membership of Management and Business Journals: A Social Network Analysis Study of the Financial Times 40', *British Journal of Management*, 21/3, 627–48.

Chan, K. C. and Fok, R. C. (2003) 'Membership on Editorial Boards and Finance Department Rankings', *Journal of Financial Research*, 26/3, 405–20.

Chan, K. C., Fung, H. G. and Lai, P. (2005) 'Membership of Editorial Boards and Rankings of Schools with International Business Orientation', *Journal of*

*International Business Studies*, 36/4, 452–69.

Cole, J. R. and Cole, S. (1973) *Social Stratification in Science*. Chicago IL: University of Chicago Press.

Frandsen, T. F. and Nicolaisen, J. (2011) 'Praise the Bridge that Carries You Over: Testing the Flattery Citation Hypothesis', *Journal of the American Society for Information Science and Technology*, 62/5, 807–18.

Garcia-Carpintero, E., Granadino, B. and Plaza, L. M. (2010) 'The Representation of Nationalities on the Editorial Boards of International Journals and the Promotion of the Scientific Output of the Same Countries', *Scientometrics*, 84/3, 799–811.

Gibbons, J. D. and Fish, M. (1991) 'Rankings of Economics Faculties and Representation on Editorial Boards of Top Journals', *Journal of Economic Education*, 22/4, 361–6.

Glanzel, W. (2008) 'Seven Myths in Bibliometrics. About Facts and Fiction in Quantitative Science Studies'. In: Kretschmer H and Havemann F (eds.) *Proceedings of WIS 2008*, pp. 1–10. Germany: Berlin.

Granger, C. W. J. (1969) 'Investigating Causal Relations by Econometric Models and Cross-Spectral Methods', *Econometrica: Journal of the Econometric Society*, 37/3, 424–38.

Inglesi-Lotz, R., Balcilar, M. and Gupta, R. (2014) 'Time-Varying Causality between Research Output and Economic Growth in Us', *Scientometrics*, 100/1, 203–16.

Kaufman, G. G. (1984) 'Rankings of Finance Department by Faculty Representation on Editorial Boards of Professional Journal: A Note', *Journal of Finance*, 39/4, 1189–95.

Laband, D. N. and Piette, M. J. (1994) 'Favoritism Versus Search for Good Papers: Empirical Evidence Regarding the Behavior of Journal Editors', *Journal of*

*Political Economy,* 102/1, 194–203.

Lee, L.-C. *et al*. (2011) 'Research Output and Economic Productivity: A Granger Causality Test', *Scientometrics*, 89/2, 465–78.

Martin, B. R. and Irvine, J. (1983) 'Assessing Basic Research: Some Partial Indicators of Scientific Progress in Radio Astronomy', *Research Policy*, 12/2, 61–90.

Medoff, M. H. (2003) 'Editorial Favoritism in Economics?', *Southern Economic Journal*, 70/2, 425–34.

Pardeck, J. T. (1992) 'Are Social Work Journal Editorial Boards Competent? Some Disquieting Data with Implications for Research on Social Work Practice', *Research on Social Work Practice*, 2/4, 487–96.

Pardeck, J. T. and Meinert, R. G. (1999) 'Scholarly Achievements of the Social Work Editorial Board and Consulting Editors: A Commentary', *Research on Social Work Practice*, 9/1, 86–91.

Prathap, G. (2006) 'Hirsch-Type Indices for Ranking Institutions' Scientific Research Output', *Current Science*, 91/11, 1439–40.

Rynes, S. L. (2006) 'Getting on Board with AMJ: Balancing Quality and Innovation in the Review Process', *Academy of Management Journal*, 49/6, 1097–102.

Sugimoto, C. R. and Cronin, B. (2012) 'Citation Gamesmanship: Testing for Evidence of Ego Bias in Peer Review', *Scientometrics*, 95/3, 851–62.

Urbancic, F. R. (2005) 'Faculty Representation of the Editorial Boards of Leading Marketing Journals: An Update of Marketing Department', *Marketing Education Review*, 15/2, 61–9.

Valle, M. and Schultz, K. (2011) 'The Etiology of Top-Tier Publications in Management', *Career Development International*, 16/3, 220–37.

Wang, J. (2013) 'Citation Time Window Choice for Research Impact Evaluation',

*Scientometrics*, 94/3, 851–72.

Willett, P. (2013) 'The Characteristics of Journal Editorial Boards in Library and Information Science', *International Journal of Knowledge Content Development & Technology*, 3/1, 5–17.

Yu, L. (2013) 'Study on Interactive Relationship of Different Sources of R&D Input and S&T Output Based on Panel Data and Panel Var Model', *Science Research Management*, 34/10, 94–102.

Zsindely, S., Schubert, A. and Braun, T. (1982) 'Editorial Gatekeeping Patterns in International Science Journals. A New Science Indicator', *Scientometrics*, 4/1, 57–68.

## Appendix A

**Questions of the E-mail Interviews for Editorial Board Members**

1. Do you think there is an influence of academic discourses in the process of reviewing manuscripts? For example, are articles having similar research topics, paradigms and philosophies to those production of editorial board members more easily recognized in academia and accepted? (The academic discourse mentioned here does not refer to academic misconduct, but emphasizes preferences and recognition in research topics, paradigms, academic perspectives and evaluation criteria.)

2. In contrast to the academic discourse noted in Question 1, is there any misuse of editorial power, for instance, to help themselves, or their universities publish unworthy articles?

3. Do editorial board members of your journal have a strong publication and citation record? What are the selection criteria involved in appointing editorial board members in your journal?

4. Is there a causal relationship between the number of editorial board members in a university and the quantity and impact of scientific output of the university? If yes, would you be able to elaborate on the dynamics of the relationship, e.g., which drives which?

5. Do you have anything else to say on the topic, that you think is especially important?

# Appendix B

## Quantile Regression Estimates (percentiles 5%–50%)

| | 5 | 10 | 15 | 20 | 25 | 30 | 35 | 40 | 45 | 50 |
|---|---|---|---|---|---|---|---|---|---|---|
| | Part A: Dependent variable (number of articles) | | | | | | | | | |
| EB | *22.462* | *26.452* | *29.762* | *31.884* | *32.530* | *33.652* | *34.667* | *35.686* | *36.647* | *38.781* |
| | *1.469* | *1.463* | *1.434* | *0.936* | *0.840* | *0.986* | *1.211* | *1.437* | *1.917* | *2.625* |
| | *0.000* | *0.000* | *0.000* | *0.000* | *0.000* | *0.000* | *0.000* | *0.000* | *0.000* | *0.000* |
| Const | *-19.923* | *-13.357* | -7.762 | 1.279 | *11.470* | *26.348* | *41.333* | *59.629* | *80.706* | *97.219* |
| | *2.711* | *3.126* | 4.242 | 3.433 | *3.952* | *6.384* | *6.512* | *8.655* | *9.700* | *11.843* |
| | *0.000* | *0.000* | 0.068 | 0.710 | *0.004* | *0.000* | *0.000* | *0.000* | *0.000* | *0.000* |
| $R^2$ | 0.212 | 0.253 | 0.280 | 0.301 | 0.315 | 0.325 | 0.333 | 0.342 | 0.347 | 0.352 |
| | Part B: Dependent variable (total number of citations) | | | | | | | | | |
| EB | *270.080* | *299.620* | *331.415* | *368.098* | *390.300* | *405.560* | *443.000* | *473.333* | *503.450* | *547.238* |
| | *13.975* | *13.044* | *12.051* | *15.570* | *11.976* | *16.675* | *22.406* | *25.360* | *26.791* | *25.087* |
| | *0.000* | *0.000* | *0.000* | *0.000* | *0.000* | *0.000* | *0.000* | *0.000* | *0.000* | *0.000* |
| Const | *-440.160* | *-299.619* | *-280.829* | *-279.098* | *-240.300* | *-159.560* | *-150.000* | -71.333 | 4.550 | 30.524 |
| | *61.621* | *24.895* | *21.213* | *29.240* | *28.413* | *38.676* | *58.369* | 71.352 | 72.787 | 84.805 |
| | *0.000* | *0.000* | *0.000* | *0.000* | *0.000* | *0.000* | *0.010* | 0.318 | 0.950 | 0.719 |
| $R^2$ | 0.210 | 0.275 | 0.312 | 0.338 | 0.359 | 0.376 | 0.391 | 0.407 | 0.423 | 0.438 |
| | Part C: Dependent variable (citations per paper) | | | | | | | | | |
| EB | *0.079* | *0.095* | *0.111* | *0.115* | *0.123* | *0.129* | *0.127* | *0.132* | *0.127* | *0.140* |
| | *0.017* | *0.016* | *0.018* | *0.016* | *0.013* | *0.013* | *0.012* | *0.012* | *0.015* | *0.017* |
| | *0.000* | *0.000* | *0.000* | *0.000* | *0.000* | *0.000* | *0.000* | *0.000* | *0.000* | *0.000* |
| Const | *4.360* | *5.021* | *5.379* | *5.845* | *6.053* | *6.284* | *6.653* | *6.875* | *7.224* | *7.430* |
| | *0.222* | *0.193* | *0.268* | *0.230* | *0.179* | *0.205* | *0.222* | *0.195* | *0.241* | *0.267* |
| | *0.000* | *0.000* | *0.000* | *0.000* | *0.000* | *0.000* | *0.000* | *0.000* | *0.000* | *0.000* |
| $R^2$ | 0.092 | 0.128 | 0.143 | 0.156 | 0.165 | 0.171 | 0.172 | 0.170 | 0.165 | 0.161 |
| | Part D: Dependent variable (h index) | | | | | | | | | |
| EB | *0.779* | *0.840* | *0.901* | *0.949* | *1.000* | *1.000* | *1.025* | *1.056* | *1.083* | *1.091* |
| | *0.076* | *0.062* | *0.060* | *0.047* | *0.051* | *0.050* | *0.047* | *0.054* | *0.055* | *0.048* |
| | *0.000* | *0.000* | *0.000* | *0.000* | *0.000* | *0.000* | *0.000* | *0.000* | *0.000* | *0.000* |
| Const | *3.221* | *5.160* | *7.099* | *8.154* | *9.000* | *11.000* | *11.975* | *12.944* | *14.167* | *14.909* |
| | *0.303* | *0.422* | *0.525* | *0.491* | *0.509* | *0.460* | *0.403* | *0.512* | *0.530* | *0.357* |
| | *0.000* | *0.000* | *0.000* | *0.000* | *0.000* | *0.000* | *0.000* | *0.000* | *0.000* | *0.000* |
| $R^2$ | 0.181 | 0.212 | 0.231 | 0.252 | 0.268 | 0.283 | 0.295 | 0.310 | 0.325 | 0.339 |

*Note.* For each of the quantiles, the following data are provided: the coefficient estimate, the standard error (1,000 bootstrapping replications), and the associated *p*-value (bold and italics denote *p*-values lower than 1%). For each quantile, we provide the pseudo $R^2$ to measure how well the data fit the estimated quantile regression models. EB represents the editorial board members, CONST is a constant, and $R^2$ is a pseudo $R^2$.

## Quantile Regression Estimates (percentiles 50%–95%)

| | 50 | 55 | 60 | 65 | 70 | 75 | 80 | 85 | 90 | 95 |
|---|---|---|---|---|---|---|---|---|---|---|
| | Partl A: Dependent variable (number of articles) | | | | | | | | | |
| EB | *38.781* | *42.250* | *46.500* | *49.364* | *51.873* | *55.125* | *60.188* | *63.680* | *70.833* | *84.706* |
| | *2.625* | *3.352* | *3.295* | *2.896* | *2.957* | *3.645* | *5.056* | *4.148* | *6.876* | *10.299* |
| | *0.000* | *0.000* | *0.000* | *0.000* | *0.000* | *0.000* | *0.000* | *0.000* | *0.000* | *0.000* |
| Const | *97.219* | *110.750* | *125.500* | *151.273* | *184.164* | *220.625* | *254.625* | *335.960* | *451.500* | *652.294* |
| | *11.843* | *12.913* | *15.120* | *15.035* | *16.028* | *17.777* | *27.071* | *31.396* | *44.562* | *76.520* |
| | *0.000* | *0.000* | *0.000* | *0.000* | *0.000* | *0.000* | *0.000* | *0.000* | *0.000* | *0.000* |
| $R^2$ | 0.352 | 0.357 | 0.362 | 0.366 | 0.369 | 0.370 | 0.368 | 0.367 | 0.367 | 0.368 |
| | Part B: Dependent variable (total number of citations) | | | | | | | | | |
| EB | *547.238* | *561.078* | *590.714* | *632.404* | *659.958* | *685.429* | *742.824* | *775.000* | *882.333* | *1109.000* |
| | *25.087* | *22.574* | *27.574* | *24.691* | *22.704* | *29.924* | *32.735* | *56.425* | *65.438* | *111.888* |
| | *0.000* | *0.000* | *0.000* | *0.000* | *0.000* | *0.000* | *0.000* | *0.000* | *0.000* | *0.000* |
| Const | 30.524 | *230.922* | *366.286* | *471.597* | *699.167* | *918.714* | *1241.158* | *1761.000* | *2443.333* | *3369.000* |
| | 84.805. | *79.542* | *88.647* | *84.611* | *103.216* | *131.468* | *152.012* | *260.389* | *311.466* | *587.190* |
| | 0.719 | *0.004* | *0.000* | *0.000* | *0.000* | *0.000* | *0.000* | *0.000* | *0.000* | *0.000* |
| $R^2$ | 0.438 | 0.454 | 0.468 | 0.482 | 0.495 | 0.507 | 0.517 | 0.524 | 0.531 | 0.536 |
| | Part C: Dependent variable (citations per paper) | | | | | | | | | |
| EB | *0.140* | *0.135* | *0.148* | *0.148* | *0.150* | *0.149* | *0.156* | *0.151* | *0.157* | *0.151* |
| | *0.017* | *0.019* | *0.018* | *0.015* | *0.012* | *0.012* | *0.011* | *0.015* | *0.024* | *0.058* |
| | *0.000* | *0.000* | *0.000* | *0.000* | *0.000* | *0.000* | *0.000* | *0.000* | *0.000* | *0.009* |
| Const | *7.430* | *7.843* | *8.141* | *8.444* | *8.761* | *9.147* | *9.485* | *10.008* | *10.715* | *12.368* |
| | *0.267* | *0.305* | *0.286* | *0.255* | *0.215* | *0.229* | *0.174* | *0.315* | *0.347* | *0.878* |
| | *0.000* | *0.000* | *0.000* | *0.000* | *0.000* | *0.000* | *0.000* | *0.000* | *0.000* | *0.000* |
| $R^2$ | 0.161 | 0.163 | 0.168 | 0.175 | 0.184 | 0.193 | 0.205 | 0.214 | 0.213 | 0.205 |
| | Part D: Dependent variable (h index) | | | | | | | | | |
| EB | *1.091* | *1.118* | *1.156* | *1.173* | *1.167* | *1.167* | *1.200* | *1.269* | *1.320* | *1.375* |
| | *0.048* | *0.060* | *0.065* | *0.058* | *0.055* | *0.061* | *0.077* | *0.068* | *0.117* | *0.188* |
| | *0.000* | *0.000* | *0.000* | *0.000* | *0.000* | *0.000* | *0.000* | *0.000* | *0.000* | *0.000* |
| Const | *14.909* | *15.882* | *16.688* | *17.827* | *18.833* | *20.667* | *21.600* | *23.154* | *26.080* | *30.625* |
| | *0.357* | *0.415* | *0.503* | *0.526* | *0.458* | *0.566* | *0.394* | *0.740* | *0.920* | *1.279* |
| | *0.000* | *0.000* | *0.000* | *0.000* | *0.000* | *0.000* | *0.000* | *0.000* | *0.000* | *0.000* |
| $R^2$ | 0.339 | 0.350 | 0.359 | 0.369 | 0.378 | 0.386 | 0.393 | 0.392 | 0.387 | 0.392 |

*Note.* For each of the quantiles, the following data are provided: the coefficient estimate, the standard error (1,000 bootstrapping replications), and the associated *p*-value (bold and italics denote *p*-values lower than 1%). For each quantile, we provide the pseudo $R^2$ to measure how well the data fit the estimated quantile regression models. EB represents the editorial board members, CONST is a constant, and $R^2$ is a pseudo $R^2$.